\documentclass{article}

\usepackage{fullpage}
\usepackage{epsf}
\usepackage{amsfonts}
\usepackage{amssymb}
\usepackage{amsmath}
\usepackage{pstricks}
\usepackage[colorlinks=true,citecolor=brown,pagebackref=true,backref=true,hyperfigures=true,hyperfootnotes=true,hyperindex=true]{hyperref}

\def\etal{{\it et.al.}\ }

\def\fig#1{Fig. (\ref{#1})}
\def\figdir{.}
\def\citeN#1{\cite{#1}}
\def\thess#1#2#3#4{$^{#3}{{\mathrm {#2}}_{#1}}^{#4}$}

\newtheorem{mydefinition}{Definition}
\newtheorem{principle}{Principle}

\begin{document}
\title{Intensional view of General Single Processor Operating Systems}
\author{Abhijat Vichare\\\texttt{abhijatv@gmail.com}}
\date{August 02, 2013}

\maketitle

\begin{abstract}  {\small  Operating   systems  are  currently  viewed
    ostensively.  As a result  they mean different things to different
    people.  The  ostensive character makes  it is hard  to understand
    OSes formally.  An intensional view can enable better formal work,
    and also  offer constructive support for  some important problems,
    e.g. OS architecture. This work  argues for an intensional view of
    operating systems.  It proposes  to overcome the current ostensive
    view by defining an OS  based on formal models of computation, and
    also  introduces  some principles.   Together  these  are used  to
    develop a framework of algorithms of single processor OS structure
    using an approach similar  to function level programming.  In this
    abridged paper we illustrate  the essential approach, discuss some
    advantages   and   limitations   and   point   out   some   future
    possibilities.}
\end{abstract}

\hrulefill
\tableofcontents
\hrulefill

\section{Introduction}
\label{sec:intro}

The purpose of  an operating system (OS; plural  OSes) is often stated
as: ``To  provide a convenient interface  to the user,  and manage the
available    hardware    resources.''    \cite{Davis:1991:OSS:531692},
\cite{Nutt:2003:OST:1208341},        \cite{Tanenbaum:2005:OSD:1076555},
\cite{Silberschatz:2007:OSC:1534584}.  This is an ostensive definition
of  an OS since  ``convenience'' and  ``interface'' are  understood by
examples and  ``management'' is specific to  hardware resources.  This
paper describes  an attempt to develop  an intensional view  of an OS.
The  approach  is  targeted  to  system  programmers,  academics  and
researchers by eliminating the ostensive view of OSes.

This work bases  OSes on basic mathematical models  of computation and
explores  the  consequences.   It  induces some  principles  based  on
practice,  and shows  that for  single  processor OSes  these offer  a
unified framework  to view the structure  of OSes.  This  work will be
useful to  diffuse the traditional  boundaries between OS  kernels and
system software.  We  do not aim for soundness  or completeness in our
approach; instead  we view  it as initial  steps toward  an eventually
sound theory of OSes.

\section{Motivation and Approach}
\label{sec:motivation:and:approach}

The  ostensive  character of  the  description  of  an OS  in  section
\ref{sec:intro} influences the theory and practice of OSes in a number
of ways.  First, different groups of people see an OS differently.  OS
programmers provide  a ``convenient'' system call  interface to system
programmers.   System programmers  provide a  ``convenient'' interface
via  techniques  like  libraries  and  system software  tools  to  the
application   programmers.     Application   programmers   provide   a
``convenient'' interface -- via the command line or graphical icons --
to the end  users.  To a system programmer  convenience could be about
the  variations supported for  resource management,  e.g.  FCFS  or RR
scheduling.  To an applications  programmer convenience could be about
the  variations   available  in  system   abstractions,  e.g.   finite
precision  arithmetic or  arbitrary precision  arithmetic.  To  an end
user   convenience  could   be  about   the   usability  abstractions,
e.g. presence or  absence of virtual desktops.  An  OS means different
things to different groups.

Secondly, the ostensive nature  of the definition affects the practice
of  development and  use of  OSes.  As  systems evolve  over  time the
conveniences may filter from the  end user level down to system level.
Thus, for  instance, we find file encryption  user utilities gradually
move  down to  the system  level where  file systems  offer encryption
capabilities.   Groups  like OS  programmers,  system programmers  and
application programmers  then need to  face the task  of cooperatively
modifying their  interfaces to  accommodate the filtering  down effect
without affecting other parts  of the overall structure.  Furthermore,
a migration duration -- during which  the new coexists with the old --
is typically offered to ease  the transition from the old architecture
to the new.   The ostensive view of  an OS does not offer  any help in
guiding the collective  modification and migration management efforts.
It  would  be  useful  to  separate  the  affected  context  from  the
unaffected ones to deal  with such architecture restructuring efforts.
It  is  also  useful  for  other  problems  like  the  development  of
virtualization strategies and formal verification efforts.  There is a
need to  examine if  there are any  unifying principles that  can help
define clear segregation and aggregation of the work across groups.

Third: An  ostensive definition of an  OS it makes it  hard to develop
any theory of  an OS.  OSes are primarily  regarded as a technological
problem.  The techniques and algorithms have evolved over the years in
response  to technological  challenges  and user  needs.  The  central
concern --  providing a convenient interface and  managing the hardware
resources -- is only  intuitively clear.  The ostensiveness percolates
down to many other critical concepts in an OS.  For example, a process
is described as an instance of a program in execution.

Researchers typically  see OSes as abstract machines.   The absence of
an intensional  definition of OSes  makes it hard to  identify general
abstract machines although subsystems  can be, and have been, studied.
Perhaps that  is why we see most  OS research is, to  quote Yates {\it
  et.    al.},   ``very   systems   oriented  and   results   driven''
\cite{Yates99i/oautomaton}.   Bergstra  and  Middelburg have  recently
pointed  out that  there appears  no insight  into what  an OS  is and
perhaps   therefore   there   may    be   no   formal   view   of   OS
\cite{DBLP:journals/corr/abs-1006-0813}.   Middelburg  in  his  survey
observes that (a)  there is ``only one more or  less abstract model by
Yates \etal \cite{Yates99i/oautomaton}'', (b) there ``does not exist a
theory based on that model'', and (c) that the OS community ``has paid
little attention  to clarifying what an  OS is and  giving motives for
introducing OSes''  \cite{DBLP:journals/corr/abs-1003-5525}.  There is
a need for an effort toward an intensional view of OSes.

Clear intensional definitions offer a  solid base to OS researchers so
that they  may define their  abstract machines precisely.   They offer
clear and constructive goals  to system programmers for building their
systems.   We hope  that the  principles we  offer support  efforts to
develop  more  intensional approaches  to  problems like  architecture
restructuring and  formal efforts for describing OS  behavior as well
as  OS  verification.  Toward  that  end  we  show that  a  systematic
framework for single processor  OSes is possible given the definitions
and principles.   The current  work describes the  essential framework
due to space limitations.

\subsection{Approaches used in Literature}
\label{sec:survey}
\label{sec:lit:review}
\label{sec:state:of:art}

\subsubsection{Origin of ostensive definitions}
\label{sec:defn:origins}

Early work  has naturally  viewed OSes ostensively  and that  view has
fossilized  over  the years.   Solntseff  succinctly  captures the  OS
evolution  in terms  of era  like the  initial ``topsy''  period where
everything not  directly produced by  the user was considered  part of
the  system \cite{Solntseff:1981:SOS:1164685.1164689}.  Denning  in an
early review paper,  offered a definition of an OS  in terms its seven
supervisory              and             control             functions
\cite{Denning:1971:TGC:356593.356595}.  An interesting footnote in the
paper  observed  that a  process  is  a  ``direct generalization''  of
procedure  in execution,  but later  work did  not build  on  it.  The
review went  on to identify five abstractions  -- programming, storage
allocation, concurrent  processes, resource allocation  and protection
-- that  could form the basis of  a ``theory'' of OS.   Dennis and van
Horn  was an  early  attempt at  figuring  out the  semantics of  OSes
\cite{Dennis:1966:PSM:365230.365252}.   A later  reprint of  the paper
defined an OS as an high level abstract machine using data abstraction
concepts  \cite{Dennis:1983:PSM:357980.357993}.  While  a lot  of work
has been done  since then, the core ideas,  definitions and algorithms
of this initial era have largely remained unchanged.

Processes were recognized  as an abstract central entity  within an OS
in the  60s and the 70s.   Denning also reviewed the  definitions of a
process and  noted that although  imprecise, they were  sufficient for
implementation purposes.  Processes were defined in various ways, some
indirectly.  Holt viewed  a process as an ``agent''  that causes state
changes  \cite{Holt:1972:DPC:850614.850627}.   Dijkstra  saw it  as  a
``sequential    automaton''    \cite{Dijkstra:1968:SLS:363095.363143}.
Dennis  and  van   Horn  described  it  as  a   ``locus  of  control''
\cite{Dennis:1966:PSM:365230.365252}.      Denning    presented    the
``instance  of  program  in  execution''  view,  and  reviewed  others
\cite{Denning:1971:TGC:356593.356595}.   Implementation techniques for
managing   memory,  processes   and  devices   were   developed.   See
\cite{Denning:1971:TGC:356593.356595}  for  a  review.  We  will  view
processes as formal computation and subsume these varied views.

The late 70s and early 80s  saw the birth of personal computing, and a
surge in innovations in hardware technologies.  A significant part was
the  hardware support  for  OS  operations, which  in  turn drove  the
systems         research.          Projects         like         Hydra
\cite{Wulf:1975:OHO:1067629.806530},                            Amoeba
\cite{Tanenbaum:1990:EAD:96267.96281},                          Medusa
\cite{Ousterhout:1980:MED:358818.358823},                       Spring
\cite{Stankovic:1989:SKN:71021.71024}            and            Accent
\cite{Rashid:1981:ACO:800216.806593}    focused   on    the   evolving
challenges of multiprocessor  and distributed systems.  Others focused
on individual sub problems in such systems.  For instance, issues like
programmability  \cite{Marques:1989:EOS:74877.74890},  monitoring  and
debugging   \cite{Tokuda:1988:RMD:68210.69222},   parallel  scheduling
\cite{Narang:2011:PDM:2007183.2007186}    and   multicore   scheduling
\cite{Shelepov:2009:HSH:1531793.1531804}  have been  addressed  in the
literature.  Distributed systems are challenging and a clear statement
of the  formal intent is  important to identify general  formal issues
among the specific problems.

\subsubsection{OS architecture issues}
\label{sec:os:archs}

Apertos  (earlier  Muse)  attempted  to  bring  in  reflectivity  (see
\cite{Smith:1984:RSL:800017.800513}) primarily motivated by the advent
of      mobile     computing     \cite{Yokote:1992:ARO:141936.141970},
\cite{Yokote:1991:MOA:122120.122122},
\cite{Yokote:1992:NSO:506378.506427}.  The central thesis of this work
is  the  separation  of  object  level  and  meta  level  abstractions
represented within the same  framework.  N\"urnberg \etal elevated the
view of  Hypermedia from a  paradigm of information organization  to a
view  of a computing  paradigm \cite{Nurnberg:1996:HOS:234828.234847}.
The separation of data, structure  and behavior is an interesting idea
in this work,  and can be seen as the  $\lambda$ calculus in disguise.
Factored  OS (FOS)  is  a recent  approach  to deal  with the  growing
complexity  of  OS  \cite{Wentzlaff:2009:FOS:1531793.1531805}.  It  is
based  on factoring  a component  (service) into  smaller ones  and is
motivated by the need for OSes  to scale up for multicore systems.  We
will offer a principle that  can guide the factorization.  In contrast
to these design  time variations of the OS  structure, approaches like
Exokernels or  SPIN investigated techniques of  mutating the structure
of the  OS at runtime.   Exokernels are the result  of re-architecting
traditional  OS  structure  to  safely expose  physical  resources  to
applications    to    allow    application   specific    customization
\cite{Engler:1995:EOS:224056.224076}.    This   is   an   example   of
redefining the interface, or the abstract machine, that an OS presents
to  applications.   The  SPIN   approach  tries  to  employ  the  good
properties  of   the  Modula-3  programming  language   to  obtain  an
extensible   system   from  a   core   set   of  extensible   services
\cite{Bershad:1995:ESP:224057.224077}.    This   safely  changes   the
interface  that an OS  presents to  an application.   While individual
issues to  focus appear distinct,  the common underlying  concerns are
about the approach to OS structuring.  Our work offers some principles
to guide such structuring efforts.

The growing  complexity of  the OS problem  has greatly  increased the
turnaround time for building experimental systems and efforts to build
such systems have reduced.  
Via the decade long K42 effort, Wisniewski \etal explore the facets of
building   a  full   OS  and   share  the   experience   and  insights
\cite{Wisniewski:2008:KLO:1341312.1341316}.  They  point out that some
important  practical  questions,  e.g.   the useful  lifetime  of  the
GNU/Linux or  Windows structures, are unanswered,  and offer plausible
reasons for  the lack of whole-OS research  efforts.  Another approach
to  deal with complexity  is to  ``go small'',  and some  efforts have
investigated OS on single user systems.  Stoy and Stratchey explored a
different control structure through the OS6 which was an early attempt
at virtual  machines.  They avoided  a job control language  through a
hierarchical  control   structure  for  system  use   (as  opposed  to
hierarchical    resource    allocation)    \cite{stoy:strachey:os6:1},
\cite{stoy:strachey:os6:2}.    Single  user   OSes   permitted  easily
removing   the  boundary  between   the  OS   and  the   user  program
\cite{Ranai:1986:DRS:382158.383030}.   Such attempts  could  have been
useful to extract the  essential abstractions, but further work needed
has  not  been  pursued.   Whole-OS  efforts, e.g.   to  describe  the
structure of an OS, are needed in addition to incremental work.

Ideas  from  programming  languages  have  been  used  to  investigate
structure and  structuring issues in  OS.  Clark discussed the  use of
upcalls      to      structure      programs      like      an      OS
\cite{Clark:1985:SSU:323627.323645}.     Kosinski   identified   eight
important issues  in expressing OS  code, e.g.  the need  for parallel
operation yet be determinate, or understandability (of OS code) in the
large  etc.  \cite{Kosinski:1973:DFL:390014.808289}.   He  developed a
data flow language based on function definition and composition, tried
to identify minimal computational function, and used these to sequence
computations.   The  Barrelfish   effort  developed  the  Filet-o-Fish
language to construct  a domain specific language (DSL)  and employ it
to generate low  level OS code \cite{Dagand:2010:FPD:1713254.1713263}.
Back \etal  use Java to explore  the OS design space,  and outline the
major technical challenges \cite{Back98javaoperating}.  They point out
that adapting  language technology to  fit into OS framework  could be
used to deal with the challenges.  Specifically, they show how garbage
collection  techniques can  be  used to  support resource  management.
Flatt  \etal demonstrate how  key OS  facilities are  obtained through
three  key extensions --  threads with  parameters, event  spaces, and
custodians    --    to    a    high   level    programming    language
\cite{Flatt:1999:PLO:317765.317793}.   They   summarize  an  important
lesson  in the  title of  their  paper --  ``Programming languages  as
operating systems''.   The bridging principle  in the current  work is
useful   to   understand   such   approaches.   Draves   \etal   apply
continuations  in  a  general   purpose  operating  system  kernel  by
redesigning  the   internal  thread  and   interprocess  communication
facilities to use continuations as  the basis for control transfer and
demonstrate    substantial   improvements   in    system   performance
\cite{Draves:1991:UCI:121133.121155}.  Using continuations allows them
to  portably  implement  new   optimization,  and  to  recast  several
optimizations found  in other operating  systems in terms of  a single
abstraction.   Kiselyov  and Shan  show  that delimited  continuations
offer  a  uniform  view  of  many  scenarios  that  arise  in  systems
programming  including an  interesting one:  a snapshot  of  a process
\cite{Kiselyov:2007:DCO:1770806.1770828}.   They implement  the zipper
file system  that explicitly  uses continuations for  multitasking and
storage.  Their  work shows  how delimited continuations  are helpful,
especially ``in conjunction with types that describe the shape of data
and  effect of  code  in detail''.   They  use such  types to  sandbox
processes,  isolate  transactions,  prevent race  conditions,  improve
scalability  to  multiple  processors,  and  obviate  the  user-kernel
boundary in hardware.

\subsubsection{Formal modeling of OSes}
\label{sec:os:formal:issues}

On the formal  side, attempts have been made to build  models of OS --
in  whole or in  part, develop  languages to  express them  and verify
them.  Yates  \etal developed a formal model  of an OS as  a system of
distributed    state   machines    \cite{Yates99i/oautomaton}.    They
investigated two  views of an OS:  a user level model  as an interface
specification,  and a kernel  level model  of the  implementation that
exposes the details  hidden by the user level  abstraction.  They show
that the  kernel level model  indeed implements the user  level model,
and hence  both are functionally  equivalent.  Another example  of the
value  of formal  work in  OS, or  its parts,  is the  graph theoretic
description          of         the          deadlocks         problem
\cite{Holt:1972:DPC:356603.356607}.  It  brought a number  of previous
deadlock algorithms  for prevention,  detection etc.  together  into a
simple  neat  structure.  On  the  verification  front, Barreto  \etal
specified process management, IPC and  file system components of an OS
kernel in  Z, uncovered errors  and inconsistencies in the  kernel and
formally verified  it by using  a mechanical theorem prover  Z-EVES on
the Z specification \cite{Barreto:2011:ASF:1945023.1945042}.

The current  work asks if an  OS can indeed  be defined intensionally.
It offers  such a view  and explores its  consequences via a  focus on
program  execution  as  the   central  concept  at  which  programming
languages  and OSes  converge.   A  good model  of  OS structure  that
identifies  the  intensional concepts  and  uses  them  to devise  the
specific  algorithms is  also  pedagogically useful  and  can help  in
activities like  OS course design \cite{Creak:2000:TOS:506117.506126}.
For  instance, traditional  memory  management and  file systems  have
redundancies  that  can  be  eliminated  to  effectively  teach  those
algorithms \cite{Esser11}.

\subsection{Approach}
\label{sec:approach}

The ostensive  nature of  the definitions used  in OS practice  can be
overcome  by connecting to  theoretical computer  science.  Algorithms
are  the  central  constructive  concept in  computer  science.   They
represent  the  part  of  our  formal  mental  concepts  that  can  be
mechanized,  i.e. concepts  for which  we can  have a  Turing machine.
Their practice involves two major aspects -- expression and execution.
Programming  languages concern  themselves with  the former,  and OSes
concern themselves with the latter.   Each uses an underlying model of
computation.  As  long as  the models are  equivalent, each can  use a
general, possibly  different, model for its purposes.   We examine the
relationship between program expression and execution to identify some
useful principles.

Once the definitions  and principles are in place, we view  an OS as a
whole.   As  an  algorithm  it  is  statically  structured  out  of  a
collection of  algorithms that together  yield a dynamic  behavior at
run  time.  The individual  component algorithms  can be  selected and
combined in a  variety of ways to realize a  given specification of an
OS.  The possible combinations form the design space of the algorithms
and define the  structure of an OS.  An OS  designer is concerned with
developing a  structure that meets the specifications.  To  obtain the
various  possible  component  algorithms  we develop  a  framework  to
describe  the structure of  an OS  based on  the ideas  of equivalence
class partitioning of resource  sets under OS operations.  However the
dynamic behavior  of processes, e.g. critical  sections and deadlocks,
is  beyond the scope  of the  current work  which restricts  itself to
rather static structural aspects of OSes.

\section{Definition and Principles}
\label{sec:definition:and:principles}

\subsection{Definition of an OS}
\label{sec:os:the:def}

Programs are  executed on,  i.e.  interpreted by,  hardware.  Hardware
implementations of  computing machines  could be regarded  as physical
realization  of some  formal  machines.   Thus we  could  see the  von
Neumann model as a physical realization of the formal Universal Turing
Machine  (UTM).  In principle,  hardware implementations  could choose
one  of the  many  equivalent formal  machines  like Turing  machines,
$\lambda$  calculus ($\lambda$),  Markov algorithms  (MA)  and partial
recursive      functions      (PRF)      \cite{Taylor:1998:MCF:275566}
\cite{Sipser:1996:ITC:524279}.   The structure  and operations  of the
chosen formal  machine form  the instruction set  of the  system.  The
instruction set  is the language to  be used to  express any algorithm
and execute it too.  Hence the formal machines must be one of the many
general models of computation.  We  call this model of computation the
\emph{low level  machine}.  On the other  hand, an end  user desires a
convenient execution system.  Such a convenient system, the \emph{high
  level machine}, must also execute any algorithm.  Hence it must also
be  one  of the  many  models of  computation.   Since  the low  level
computation model need  not be the same as  the high level computation
model,  we introduce  an algorithm  over  the low  level machine  that
transforms the  system into  the high level  machine, and call  it the
\emph{Operating System}.  An intensional definition of an OS is:

\begin{mydefinition}
  \label{basic:os:def}
  An \emph{Operating System} is  an algorithm that implements a Turing
  complete  model of computation,  called the  high level  machine, in
  terms  of  another  Turing  complete  model, called  the  low  level
  machine.
\end{mydefinition}

\begin{figure}[h]
  \centering
  \epsfxsize=.5\linewidth
  \epsffile{\figdir/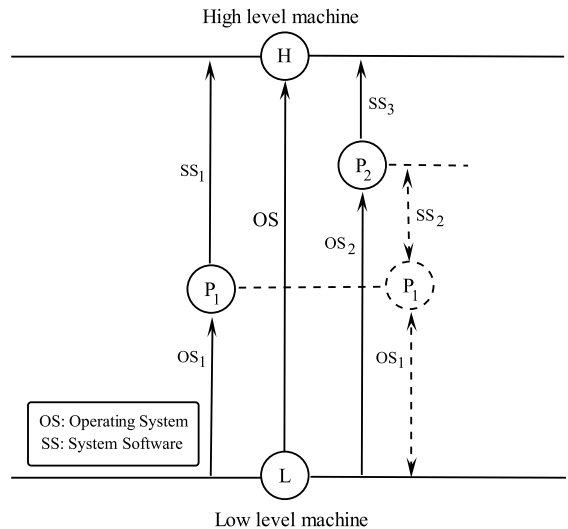}
  \caption{Pictorial  view  of  an  OS as  conceptually  described  by
    definition  \ref{basic:os:def}  and   its  practice  (see  section
    \ref{sec:os:practice}).  High level  languages P$_1$ and P$_2$ are
    depicted to suggest two different levels of abstraction.}
  \label{fig:conceptual:os}
\end{figure}

The notion of an OS is now  based on the theory of computation.  It is
an algorithmic  proof of  equivalence between the  low level  and high
level Turing  complete models of computation.   The structural aspects
of  an OS  are as  a bridging  algorithm between  a pair  of universal
Turing complete  machines.  The operational  aspects of an OS  are all
the possible  ways in which the  high level machine  may interpret its
programs.   As an  algorithm, an  OS  has to  realize its  operational
aspects  using the  low level  machine.  Fig.(\ref{fig:conceptual:os})
captures this  intensional definition  \ref{basic:os:def} of an  OS as
the algorithm between a low level machine $L$ and a high level machine
$H$.
Further,  this definition  allows  us to  introduce another  ostensive
concept --  the \emph{abstraction gap} -- that  captures the intuitive
difference between the  two models of computation.  The  purpose of an
OS is to bridge this abstraction gap.

Another  interesting description  of an  OS emerges  from  the program
expression point  of view.  As  the high level  machine of an OS  is a
universal  machine,  the  OS  accepts  descriptions  of  one  or  more
machines, i.e.  programs, and executes  them.  The OS and its programs
must be fully  decoupled from each other.  Programs  may be defined at
any time, any where and in any way possible.  Once defined they may be
executed  any time,  any where  and in  any way  possible.  An  OS may
therefore be described as a  program whose one or more subprograms are
fully  decoupled from  itself.   In  this sense,  every  program is  a
delimited    continuation    as    suggested   by    Kiselyov    \etal
\cite{Kiselyov:2007:DCO:1770806.1770828}.

\subsubsection{Correspondence with practice}
\label{sec:os:practice}

The low level machine language  may not always be suitable for program
expression  purposes.   In practice   we  often  introduce  another
computation model, usually Turing complete, that is used to design the
program expression system.  We call  the instruction set of this model
of computation as the \emph{high level language (HLL)}.  More than one
HLLs may  exist in a  system.  Fig.(\ref{fig:conceptual:os}) expresses
this idea and $P_1$ and $P_2$ may denote two HLLs.  For example, $P_1$
may be the HLL defined by the system services of typical Unix style OS
kernel,  $P_2$  could  be  the   HLL  defined  by  a  typical  program
development HLL, say C.

Fig.(\ref{fig:conceptual:os})  shows that  an  OS as  a  concept is  a
bridge between a low level machine and another high level machine.  It
may  be regarded  as  a  whole entity  or  may be  composed  of an  OS
component and a system software  (SS) component.  As shown for machine
$P_2$,  the $OS_2$ may  be composed  of $OS_1$  and a  system software
component  for expression  $SS_2$.  Note  that $OS_2$  need not  be so
composed and  can stand alone.  Additional system  software, $SS_3$ --
usually the software tools, can be  used to reach the level of machine
$H$.  Whether the high level machines are $H$ or $P_1$ or $P_2$, an OS
is essentially the bridge between them and the low level machine $L$.

\label{notation}
\textbf{Notation:} Let  \thess{I}{S}{L}{H} denote that  a concept $S$,
e.g. an  OS or system software SS  or language HLL, be  expressed in a
language $I$ to raise a low  level machine {L} to a high level machine
$H$.  Thus  \thess{\_}{OS}{L}{H} is an OS expressed  in some language,
denoted by $\_$, that raises the low level machine $L$ to a high level
machine  $H$.   Similarly,  \thess{C}{SS}{C}{\lambda}  denotes  system
software expressed in C to  obtain a $\lambda$ calculus machine over a
given  C  machine.   The  notation  is  adapted  from  the  T  diagram
introduced  by Bratman  in \cite{bratman}  and  used by  Aho \etal  in
\cite{aho-sethi-ullman}.

Designing  HLLs, i.e.  program  expression systems,  involves defining
frameworks  over  a  chosen  model  of  computation.   Frameworks  may
introduce arbitrary  rules over the model to  ease program expression,
but that  may not  affect the computation  power of the  model chosen.
For instance,  the C programming  language introduces a rule  that the
occurrence of a  function call at runtime directly  corresponds to its
syntactic occurrence  in the  source code.  This  does not  affect the
ability to  express any computation  sequence but may  make expressing
some of them inconvenient.  For instance, consider the \texttt{signal}
abstraction to express response to events.  The signal handler call is
tied up to  event occurrence and not with  the syntactic occurrence of
the  handler  in  the  expressed  C  programs.   $P_1$  and  $P_2$  in
Fig.(\ref{fig:conceptual:os})  are often  frameworks over  some formal
machines.

In general  any OS, \thess{\_}{OS}{L}{H}  or \thess{\_}{OS}{L}{P_1} or
\thess{\_}{OS}{L}{P_2},  may  not   restrict  program  definition  and
execution in any way.  However  the generality of the algorithms in an
OS that manipulate the state and control flows determine the extent of
the  restrictions on program  execution.  An  OS that  uses coroutines
style  approach to  control flow  and state  manipulation can  at best
support non-preemptive multitasking.

\subsection{Procedures and Processes}
\label{sec:process:and:file}

A  process  is  often  defined  as  ``an  instance  of  a  program  in
execution'';     see    \cite{Denning:1971:TGC:356593.356595}.     The
definition  is   esoteric  since  a  process   is  understood  \emph{a
  posteriori}, i.e. after one has  some experience with the ideas of a
``program  instance''  and its  ``execution''.   Dennis  and van  Horn
define a process in an \emph{a priori}, but more abstract, way as: ``A
process is  that abstract entity which moves  through the instructions
of  a  procedure  as  the  procedure is  executed  by  a  processor.''
\cite{Dennis:1966:PSM:365230.365252}.  Both these definitions are also
contingent ones and do not clearly capture the mechanical aspects that
are  required  for formal  work  as  well  as for  implementation.   A
mechanical view  of process would  capture the operational  meaning of
``instantiating a program'' and ``executing it''.

To obtain a mechanical view we use the intuition behind interpretation
as    formally    defined   in    elementary    proof   theory    (see
\citeN{Gries:1993:LAD:161182},     for     example).      Intuitively,
interpretation is  the ability to extract  the value bound  to a given
symbol.  A  computing machine interprets  the symbols of  its programs
given  a  set  of  values  bound  to them.   These  bindings  are  the
\emph{state}, or  an \emph{environment} or  \emph{context}.  A process
is,       therefore,      the       formal       computation      (see
\cite{Sipser:1996:ITC:524279}  \cite{Taylor:1998:MCF:275566} etc.)  by
a  computing  machine when  a  set  of symbols  of  a  program and  an
associated context are presented to the interpreter.
\begin{mydefinition}
  \label{def:process}
  Given a program and  an associated environment for interpretation, a
  \emph{process} is the formal computation by a formal machine, e.g. a
  Turing machine.
\end{mydefinition}

We will  refer to  the set  of symbols --  program and  the associated
context  -- as  a \emph{procedure}.   A  procedure is  a process  when
interpreted  by a  machine.  A  process induces  state changes  in the
system and is  an active entity \cite{Denning:1971:TGC:356593.356595},
\cite{Holt:1972:DPC:850614.850627}.  A  procedure is a  passive entity
when not interpreted  by a machine.  When passive we  will refer to it
as a  \emph{file}.  A procedure is  a structural aspect of  an OS, and
forms the basic  unit of manipulation by the algorithms  of an OS.  It
is  the  final goal  of  program expression  while  a  process is  its
specific formal  computation.  This distinction  between the procedure
and its process helps to  avoid the contingent aspects.  ``Executing a
program'' is the formal computation of a procedure.  ``Instantiation a
program''  is  creating  the  procedure,  i.e.  the  program  and  its
associated interpretation context.  It  gives a clear algorithmic goal
for system programmers who devise system software, e.g.  loaders.  For
them,  a procedure  is the  structural  entity that  must be  switched
between active and passive states.

The procedure in the above definition corresponds exactly to the input
to  a Universal Turing  Machine (UTM).   This notion  of a  process as
formal   computation  is   also  invariant   over  all   the  computer
organization and  computation organization strategies,  whether single
processor  machines  or  distributed  systems.   It  only  requires  a
suitable underlying formal model of computation.

\subsection{Some useful principles}
\label{sec:useful:principles}

Principles capture invariants of a problem that can be used to develop
a structure, i.e.   the steps and their permutations  that satisfy the
specification,  of the  solution.  The  questions that  help capturing
them for the OS problem as given by definition \ref{basic:os:def} are:
(a) What invariants can be induced?, (b) How can we identify the steps
required by an OS to bridge  two models of computation?, and (c) Given
the steps what determines their correct arrangement?

\subsubsection{First classness}
\label{sec:first:classness}

Our definition \ref{basic:os:def} offers  an invariant.  An OS must be
free  to execute  a  set of  one  or more  procedures  with the  least
restrictions,  i.e.   any where,  any  way  and  any time.   For  each
procedure it must define two  states: active (or running), and passive
(or ready to run).  To  realize execution with the least restrictions,
procedures must  be switched between  these two states in  any desired
manner,  i.e.    procedures  must  be   \emph{first  class}  entities.
Procedures are  first class since as files  they can be a  member of a
set  used to  select for  scheduling.  They  are also  a  parameter to
decision  functions  that fix  the  schedule,  or  a return  value  of
selection  functions  that  pick  a  file to  schedule.   They  become
processes  when subject  to  interpretation.  The  first classness  of
procedures must be an invariant property.
\begin{principle}[First classness]
  \label{first:classness:principle}
  Procedures in an OS must be first class.
\end{principle}

First classness  of procedures is necessary  to allow an  OS to evolve
their execution  in any pattern.  Now the  possible execution patterns
are the  number of  distinct ways  in which an  OS may  switch between
active and passive states of each procedure in the set.

\subsubsection{The Bridging Principle}
\label{sec:os:and:prog:lang}

To realize, i.e.  design and implement,  an OS as a bridge between the
two levels of machines, a designer usually introduces intermediate but
intuitively  clear  levels  of  abstraction.   Each  of  these  levels
preferably captures one step  towards the eventual high level machine.
Each  intermediate  level must  also  be  Turing  complete since  each
subsequent level must be able to express all algorithms.  The question
is: how do we identify these levels?

We now add  a program expression system that is used  to express an OS
between a  low level machine and  a high level  machine.  This system,
although a framework as a  whole, is defined by some underlying formal
computation  machine.  We  will  call this  framework  as the  \emph{E
  machine},  the expression machine.   There are  three cases:  (a) $L
\not = E \not = H$, e.g.  \thess{C}{OS}{\lambda}{MA} (b) $L = E \not =
H$, e.g.   \thess{C}{OS}{TM}{\lambda} and  (c) $L \not  = E =  H$ e.g.
\thess{C}{OS}{\lambda}{TM}.\footnote{The     notation     (see    page
  \pageref{notation}):  \thess{C}{OS}{\lambda}{MA} denotes an  OS that
  realizes a  Markov Algorithms (MA)  based high level machine  over a
  $\lambda$  calculus based  low level  machine, with  the development
  hosted on an Turing machine based HLL expression system, say C.}

In each  of the three  cases an OS  has to necessarily bridge  any gap
between the  $E$ machine and the  high level machine.   The details of
the three cases differ in the techniques to bridge the gap between the
low level machine and the $E$  machine so as to achieve execution with
least possible  restrictions.  In case  these two are  different, then
the OS has  to first implement the low  level machine completely using
the $E$ machine.  In the case when  the $E$ machine is the same as the
low level machine the OS  may have to implement algorithms to overcome
the restrictions by parts of the  HLL framework.  Thus, given C as the
chosen  HLL an  OS  has to  overcome  the restriction  of stack  based
activations  and  implement more  general  activation mechanisms  like
event  based activation  (i.e interrupts)  or  multiplexed activations
across processes for  system calls.  In the case  when the $E$ machine
is the same as the high level machine the OS may have little to bridge
except algorithms  to overcome  the restrictions by  parts of  the HLL
framework.

Given  the  chosen level  of  program  expression  the three  bridging
possibilities yield  the bridging principle below.   It identifies the
origin and the nature of  the abstraction gap alluded to by definition
\ref{basic:os:def}.  It  provides a mechanism to  obtain the necessary
steps to reach the high level model defined for the OS.
\begin{principle}[The Bridging Principle]
  \label{bridging:principle}
  An OS must necessarily bridge any difference between the $E$ machine
  and the  high level  machine, and it  must necessarily  overcome any
  restrictions  introduced  by the  $E$  machine  over  the low  level
  machine.
\end{principle}

To illustrate  the bridging principle, consider the  second case above
\thess{C}{OS}{TM}{\lambda}  where we have  the C  $E$ machine  used to
implement an  OS that bridges  a von Neumann  low level machine  and a
$\lambda$   calculus   high   level   machine.   With   reference   to
\fig{fig:conceptual:os}, we  have $L$ as the von  Neumann machine, $H$
as the $\lambda$  calculus based machine, and $P_2$ as  the C HLL that
realizes the C semantics.  To  reach the full $\lambda$ machine level,
the  $SS_3$  component  is   necessary.   It  includes  realizing  the
abstraction  that  are  missing  from   the  C  HLL  as  well  as  the
abstractions that need to be  generalized to overcome the framework in
C.  C does  not have automatic memory management,  and its stack based
activation needs  to be generalized.   On Unix like systems  $SS_3$ is
often  a set  of  programs  and libraries  that  provide a  functional
interface that  hides memory  details and also  introduces composition
capabilities.   The creation of  generalized activation  mechanisms is
moved into the kernel, e.g.  per process u area to realize multiplexed
system calls.

\begin{figure}[t]
  \centering
  \epsfxsize=.5\textwidth
  \epsffile{\figdir/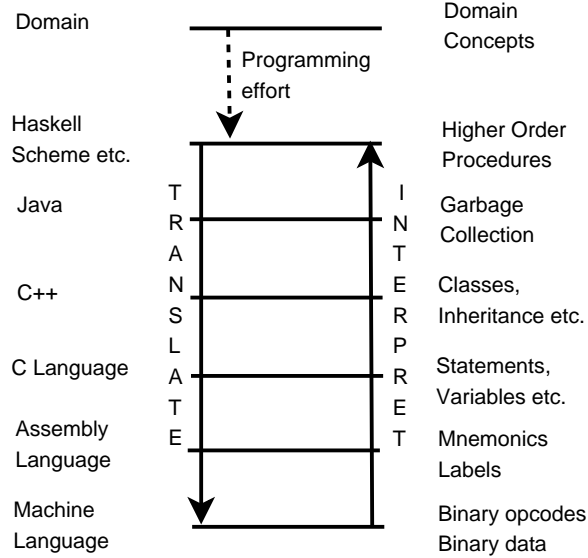}
  \caption[Abstraction    of   expression   levels    in   programming
  languages]{Abstraction of expression  levels in programming language
    design gradually evolving from  von Neumann machines to functional
    machines.  The figure shows typical lowest and highest levels.}
  \label{fig:prog:lang:abstractions}
\end{figure}

To  identify  the   intermediate  abstraction  levels,  including  the
limitations of the  framework, for \thess{C}{OS}{TM}{\lambda}, we look
at     the    abstraction     levels     of    program     expression.
\fig{fig:prog:lang:abstractions}  arranges them  to correspond  to the
intuitively increasing  sense of  the abstraction levels.   The figure
makes the point that the arrangement captures a gradual transformation
from  a  Turing  machines  based computation  model  towards  eventual
functional one.
On the  right is a  set of  levels of expression  and on the  left are
example languages that support them.  They are arbitrarily arranged to
appear to increase upwards from  the bottom.  One may either translate
expressed programs down to the  low level interpreter machine as shown
by the arrow  on the left or raise the interpreting  machine up to the
level of  the expressed programs as  shown by the arrow  on the right.
They gradually achieve the functional programming model of computation
by  creating  better name  spaces  (e.g.   mnemonics, variable  names,
function  names),   state  spaces  (e.g.    types),  automatic  memory
management  (e.g.  garbage  collection),  and interleaving  executions
(e.g.  closures and continuations).

The  preceding  discussion   suggests  that  techniques  of  computing
closures  and continuations generally  must be  borrowed into  the our
specific OS  example problem  since the C  HLL based  expression model
supports only stack based activations.  Also the algorithms that would
eventually  create  implicit memory  for  the  high  level model  from
explicit primary  memory of  the low level  model are  also necessary.
The bridging principle  thus helps an OS to  identify the intermediate
levels necessary  to bridge the  two models of  computation.  However,
since     the    idea    of     ordering    the     abstractions    in
\fig{fig:prog:lang:abstractions} is  currently intuitive, the bridging
principle is ostensive and contingent.

\subsubsection{Binding and Interpretation}
\label{sec:functional:end}

Having  identified the intermediate  steps between  the low  level and
high level machines, we organize them in a sequence that satisfies the
specification of an  OS.  The design space is formed  by the number of
allowed ways  in which the  given set of  steps may be  permuted.  The
binding principle  that follows  is the decision  rule to decide  if a
given permutation of the identified steps is acceptable or not.

Given that interpretation is the ability to extract the value bound to
a given symbol,  the central construct required is  binding symbols to
their  values.  The interpretation  system is  discrete and  takes one
symbol at a time for interpretation.  For successful interpretation we
need the following, stated as the principle of binding:

\begin{principle}[The Binding Principle]
  \label{binding:principle}
  The binding of a value to a symbol may be established at any instant
  before the instant of interpretation.
\end{principle}

\begin{figure}[ht]
  \centering
  \epsfxsize=.5\linewidth
  \epsffile{\figdir/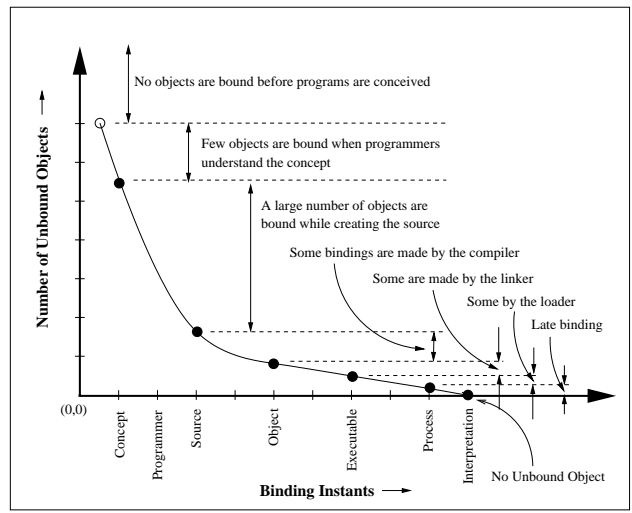}
  \caption[Temporal sequence  of bindings]{Temporal sequence  in which
    bindings are gradually established by the translation technique.}
  \label{fig:translations:binding:seq}
\end{figure}

The bridging  principle helps  introduce the intermediate  steps which
can be  permuted in any way.   The binding principle is  the rule that
defines acceptable permutations.  Together  they can yield an approach
similar  to  the def-use  chains  in  data  flow analysis  that  yield
compiler   optimizations   like   constant  propagation   and   common
subexpression elimination.  Since  interpreters are acceptor machines,
the  binding principle  captures  the need  to  establish the  binding
before  use.  Given  a  sequence of  symbols  for interpretation,  the
principle forms  the base of  the tool chain approach  of establishing
binding  analogous  to  compilation  phases.   Bindings  that  can  be
computed earlier  are moved to the  earlier phases of  the tool chain,
e.g.   libraries.   Fig.(\ref{fig:translations:binding:seq})  captures
this idea for a typical compilation sequence.

Given that  the bridging principle helps to  identify the necessities,
the binding  principle helps to identify the  correct possibilities of
placing them.  It helps to identify the design choices required to map
the low level machine to  the high level machine.  For instance, since
framing  the main  memory, i.e.  binding frame  numbers to  fixed size
memory regions, is independent of  dividing a program into pages (i.e.
binding  page numbers  to fixed  sized  program chunks),  both can  be
established independently and anytime before the page table that binds
pages to frames  is built.  As another example,  the file system layer
within a monolithic  kernel can be viewed as a set  of bindings of the
file  system abstractions  to  memory devices  moved  into the  kernel
space.  On the other hand,  if the device abstractions ensure disjoint
regions then the file system abstractions for each region may be moved
into user space too.  The  impact of this principle for conceiving and
building system software will be discussed in another work.

\section{Resource Management Algorithms}
\label{sec:resource:mgmt:algos}

The detailed  version of  this paper works  out the  equivalence class
partitioning idea, and illustrates  a complete paging and segmentation
based memory  management algorithm.  It is based  on partial recursive
functions model of computation and uses a Lisp like notation to better
describe  the  formal  structure.   The description  below  sacrifices
little  accuracy for  brevity of  illustration of  the  approach.  For
example, it allows signature inconsistency in composition.

For  an OS  a procedure  $p$ is  characterized by  two  parameters: $p
\equiv p(s, t)$, where $s$ is the memory size and $t$ is the CPU time.
An OS is  given a (assumed finite) set of  procedures, $P = \{p_1(s_1,
t_1),  p_2(s_2, t_2),  \ldots,  p_n(s_n, t_n)\}$,  of $n$  procedures.
Memory resources are sets of  finite and reusable units while CPU time
is infinite and not reusable units.  The set of memory units, $M$, and
the set of CPU time instants,  $T$, are examples of resource sets that
we  generically denote as  $R$.  In  general, resources  are countable
sets  of units,  $R =  \{r_1, r_2,  \ldots\}$.  To  each element  of a
resource set $R$ we define  a function $addr$ that associates a unique
natural  number  called  the  \emph{address}  of the  element  of  the
resource  set.   $addr$ can  be  used to  order  the  elements of  $R$
monotonically,  and we  assume  that  they have  been  so ordered.   A
contiguous subset $S \subseteq R$ is specified by two addresses -- the
start address of the subset and the end address of the subset.

The structural  algorithms of an  OS are about managing  the resources
like memory and CPU time.   We see these algorithms as operations that
induce an equivalence class  partitioning of the resources.  To obtain
specific  algorithms, we  look  at  the different  sets  (e.g. $P$  of
procedures or  $R$ of resource units), their  structure, and different
possibilities  of binding  resource sets  to specific  resources.

The resource allocation problem is: given the required quantity $q$ of
the resource to  extract a non-intersecting subset of  $S \subseteq R$
of contiguous elements  of $R$ such that $\#S  \equiv (addr(r_{end}) -
addr(r_{start}))  = q$.   The  required quantity  $q$  is obtained  by
selecting  an  element  from  $P$,  and projecting  out  the  required
parameter.   Thus if  $R$ denotes  the  memory, then  $q$ denotes  the
projection of  the size parameter  $s_i$ of the $i^{th}$  procedure in
$P$.  The \emph{select} operation that  selects an element of a set is
any algorithm that  returns a member of the set.   A simple example of
\emph{select} is  the projection function that given  a natural number
$i: 1  \le i \le  \#R$ returns the  $i^{th}$ element of set  $R$.  The
elements  of a  set $R$  may be  reorganized using  an \emph{organize}
operation that  rearranges the  elements of $R$.   The \emph{organize}
operation  operates  on $R$  before  the  \emph{select} operation.   A
simple  example of  \emph{organize} operation  is  the \emph{identity}
function that given a natural number  $i: 1 \le i \le \#R$ returns the
$i^{th}$  element  of set  $R$.   Another  example of  \emph{organize}
operation is  the \emph{fixed size partitioning}  algorithm that first
organizes the  resource units into  fixed sized allocation  units, and
then selects  an allocation unit  according to an $addr$  defined over
them.   Composing  the   specific  \emph{select}  and  \emph{organize}
operations on  some resource $R$  yields a family of  algorithms.  For
instance, the first-come-first-served  algorithm is simply: $identity\
(identity\ (R))$,  where we have used the  \emph{identity} function as
instances  of   both,  the  \emph{select}   and  the  \emph{organize},
operations.  The  buddy allocation algorithm  may be considered  to be
composed  of a  tree  traversal  based selection  that  operates on  a
resource set organized into a binary tree.

The set $P$  may also be subject to  reorganization independent of any
organization   of   the   set   $R$   of   resource   elements.    The
shortest-job-size-first  algorithm is:  $identity\ (sort\  (P, s_i))$,
where $s_i  \equiv project\  (1, project\ (i,  P))$, projects  out the
first  parameter,  size, of  the  $i^{th}$  procedure  from $P$.   The
shortest-job-time-first  algorithm is:  $identity\ (sort\  (P, t_i))$,
where $t_i  \equiv project\  (2, project\ (i,  P))$, projects  out the
second parameter, CPU  time, of the $i^{th}$ procedure  from $P$.  The
\emph{sort}   operation  reorganizes   $R$  and   the  \emph{identity}
operation  selects  an element  from  the  reorganized  set.  If  each
element of $P$ is  arbitrarily associated with an externally specified
natural number, the priority of  the procedure, we have priority based
procedure  selection.  Since  a  high priority  process  must also  be
allocated  memory  resource  (i.e.   $R  \equiv M$),  we  compose  the
manipulation of $P$  and the manipulation of the  memory resource $M$.
If  we  use buddy  allocator  for manipulating  $M$,  we  have a  high
priority  procedure  allocated   using  buddy  methods.   The  binding
principle demands that the  priority selection mechanism and the buddy
organization  mechanism  may  be   established  in  any  order  before
combining them.

Until this  point we have  assumed that procedures are  indivisible in
the  sense that  their  resource  needs are  either  to be  completely
satisfied or not at all.  However, it is possible to partially satisfy
the resource needs by dividing  the needs into smaller chunks.  If the
chunking is  variable sized, e.g.   due to cognizance of  the internal
structure  of  the  procedure,  we have  segmentation  algorithms  for
memory,  and I/O-bound--CPU-bound  algorithms  for CPU  time.  If  the
chunking disregards any internal structure and partitions the resource
needs into equal chunks we  have paging algorithms for memory resource
and  round robin algorithm  for CPU  time resource.   Additionally, we
have  assumed that  the cardinality  of sets,  like $P$,  is constant.
Using a  countably infinite  set $P$ yields  a more realistic  OS that
keeps executing  processes as  they are fired.   Also the size  of the
elements  of the  sets, e.g.  a procedure  in $P$,  is  also constant.
Varying  sizes of  procedures allows  mechanisms like  plugins.  These
variations must be designed in accordance with the binding principle.

Another implicit assumption is that  a specification of a set, say $R$
of  some resource, is  bound to  the corresponding  physical resource.
This can  be relaxed to yield  an interesting possibility.   If $R$ is
instead bound to another set  $R^\prime$, and $R^\prime$ is then bound
to the actual resource,  we have \emph{resource virtualization}.  This
is  used in  segmentation and  paging  methods in  OSes.  The  primary
memory  resource  specification is  bound  to  a  set $M$  over  which
segmentation is  defined.  $M$ is  then bound to the  physical primary
memory $M^\prime$ over which  paging is defined, with $\#M^\prime \not
= \#M$ as a possibility.

Finally, we have implicitly assumed  that the subsets of resources are
always  disjoint.    The  disjointness   will  always  hold   for  the
non-reusable CPU time resource  on single processor computing systems.
However, for  the reusable memory  resource this disjointness  may not
hold.  If two distinct subsets  have a non-empty intersection, we have
resource  sharing, and  write  accesses  to the  shared  area must  be
temporally sequenced.   This is the critical section  problem, and its
temporal  nature  cannot  be  adequately captured  by  our  structural
approach.

To  preserve  space,  we  have  presented the  essential  approach  to
resource management algorithms using subsets.  The details examine the
various  implicit assumptions as  indicated above,  and relax  them to
obtain the framework.  Also included are other properties of resources
like  finiteness  and  reusability   (e.g.   memories)  that  lead  to
deallocation algorithms and non primary property of secondary memories
for swapping techniques.  Swapping is  viewed as a mechanism to extend
the primary memory using a fraction of the secondary memory given that
procedures  must be  swapped  back to  primary  memory for  execution.
Security  issues are  also  not  shown since  they  are influenced  by
external  environment   of  the  computing  system   that  drives  the
definition of  resource subsets.  However,  they may be  considered as
inducing  equivalence class partitioning  based on  externally defined
criteria like ownership.

Our approach of describing resource  management views the problem as a
binding problem between  a procedures set $P$ and  a resource set $R$.
Resources   management   operations   induce  an   equivalence   class
partitioning of the sets, and the variations in binding the members of
the two emerge  from the manipulation of the  sets and their elements.
Various   algorithms  that   perform  the   generic   operations  like
\emph{select} and \emph{organize}, and their compositions according to
the  principles  in   section  \ref{sec:useful:principles}  yield  the
various specific  structural algorithms of an OS.   This approach thus
unifies  the algorithms  into  a systematic  framework.  However,  our
approach  does not  yield  algorithms that  need  to consider  process
behavior, e.g.  deadlocks and synchronization.

\section{Application to a Unix like OS}
\label{sec:application}

So  far we  have  intensionally  defined an  OS,  induced some  useful
principles,  and  formally  described  the resource  algorithms  using
subsets.  We  now discuss their impact  on the architecture  of a Unix
like    OS     whose    kernel    is    as     described    by    Bach
\cite{Bach:1986:DUO:8570}.  An  example is the GNU/Linux  OS, which is
made  up of  the kernel  and  other system  software.  The  definition
\ref{basic:os:def} requires  a well  defined high level  machine.  The
$\lambda$ calculus model  is a good high level machine  for an OS.  It
is an intuitively simple view for  an end user since it strips off all
the unnecessary details.  After all, an algorithm needs the ability to
identify its  distinct component objects,  prescriptions that describe
transformation of some objects into  others, and an ability to apply a
prescription to  the components  of the algorithm.   The rules  of the
$\lambda$  calculus  capture  this  intuition.   A  computer  user  is
typically  concerned  with  obtaining  the results  of  applying  some
transformation  on  some  objects.   Technological  details  like  the
representation of the components,  their storage and retrieval, or the
variety of ways in which they may be made to interact via applications
are  not relevant to  the end  user.  The  $\lambda$ calculus  view of
computation can be  the candidate high level machine  if the technical
details  are   hidden  away   through  techniques  like   file  format
standardization and  implicit memory management.  The  user level view
and  the  kernel level  views  in  Yates  \etal respectively  can  now
correspond to the high level almost functional machine (the user view)
and   a    low   level   imperative   machine    (the   kernel   view)
\cite{Yates99i/oautomaton}.

The GNU/Linux  is built  over low level  Turing machine  like hardware
with an empirically  defined higher level.  It is  an almost $\lambda$
calculus  type  machine  with  sophisticated naming  capabilities  via
icons, application capabilities  via drag-and-drop techniques, and the
compositional capabilities of the command line.  It still has some way
to  go to  reach full  capabilities of  the functional  approach.  For
instance, anonymous  $\lambda$ expressions  are still to  be realized,
and still  more simply not all programs  return values  called ``exit
codes'' to the OS.

Despite  these  shortcomings we  argue  that  GNU/Linux  aims for  the
$\lambda$  calculus model as  the high  level computation  model.  The
$\lambda$   calculus  requires   three  abilities:   naming,  function
abstraction, and application.  As  a candidate high level machine for
an OS, the reductions are implicit within the OS.  At the low level we
are given  an infinite  countable tape --  the primary memory,  a head
that captures the  state transition abilities -- the  CPU ISA, and the
basic read-write operations on the tape.

A name  binds a symbol to  an object in the  system.  Within GNU/Linux
procedures are  named using process  identifiers, and files  are named
using  strings.   Other  ways  of  naming  files  are  through  memory
addresses  (e.g.  register names,  memory location  values, pointers),
I/O ports  (hardware or software  enumerated), and device  names.  The
abstraction  level of  names is  elevated by  using file  systems that
organize primitive  names into path  names that are bound  to objects.
In the  primary memory, files  are named using natural  numbers called
file descriptors.

A function  application is expressed simply as  program with arguments
for operation, or even as  a drag-and-drop operation.  In response, an
OS  loads the program  into primary  memory as  a file,  completes any
contextual  bindings required,  and obtains  a procedure  that  can be
scheduled.  This is the conventional process.

To run a function application, a CPU time schedule must be determined.
An OS  has a set of  one or more applications  to run, and  a CPU time
allocation strategy.   To switch between the  function applications an
OS computes the closure to capture the data state and the continuation
to capture  the control state.  These may  additionally be manipulated
too.  For example, an  \texttt{exec()} call overwrites the closure and
continuation of the parent with new ones.

Most  current OSes  distinguish between  ``executables''  and ``(data)
files'', particularly for security and trust reasons.  First classness
shows that such  distinction is not fundamental to  OSes.  The binding
principle is used to determine  the boundaries between the kernel, the
system  libraries, the  system software  and application  software.  A
monolithic kernel can  be viewed as a single  set of bindings obtained
by moving as many bindings as possible as early as possible into their
expression.   Exokernels support  reshaping the  bindings  at runtime.
Microkernels further  refine the single set of  bindings into multiple
sets.   System  libraries  may   enhance  the  kernel  bindings,  e.g.
\texttt{socket} abstractions for IPC, and language libraries may adapt
bindings, e.g.  \texttt{fopen()}  over \texttt{open()}.  Libraries can
also extend the abstraction level of the OS, e.g.  garbage collectors.
While  the  bridging principle  can  identify  the intermediates,  the
binding principle defines the allowed permutations.

An  OS like the  GNU/Linux can  be viewed  as an  yet to  be completed
program  to create  a  functional  style high  level  machine over  an
imperative low level machine.  It  is difficult to express and execute
higher order procedures, anonymous procedures etc.  Lazy evaluation is
also partial, e.g. Unix pipes.

\section{Conclusions and future work}
\label{sec:conclusions}

An intensional  definition of an  OS and procedures, and  some induced
principles  is the main  core of  this work.   This attempt  yields an
insight into  the connection between the formal  models of computation
and OS practice.  We separate the structural and behavioral aspects of
OSes,  and  handle  the   structural  aspects  via  equivalence  class
partitioning  of  resource  sets.   We present  some  principles  that
capture general  properties, help identify the  intermediate steps and
define their  correct arrangement during the  bridging process.  While
some principles  are directly useful in devising  algorithms, some are
still ostensive.  Given a well defined  purpose of a Unix like OS, our
approach offers a high level understanding and can also point out some
next technical directions.

The  behavioral aspects  of  an  OS need  a  suitable process  algebra
formulation.  Other  future work involves carrying  out better studies
of the abstract machines in an  OS via techniques like ASM or abstract
interpretation.    In  principle,   adapting  some   program  analysis
formalisms  (e.g.   \cite{Atkinson:1996:DWA:227726.227732})  for  OSes
could  be   useful  for  exokernels,   microkernels  and  architecture
restructuring,  but is a  significant challenge.   Another interesting
possibility is  a more  declarative view of  an OS with  potential for
generation  of  its  imperative   view  and  for  formal  verification
problems.  Parallel computing systems  cannot be handled at the moment
due  to  the  lack  of   a  good  model  to  capture  variations  from
co-processing to cluster computing to fully distributed computing.

We also envisage  a more esoteric future work.  The  world view of the
physical  sciences  expresses   natural  laws  through  Lagrangian  or
Hamiltonian  formulations that  are  based on  an extremum  principle.
Given a  universal machine view of  an OS, we may  view computation as
made  up of  a  ``kinetic''  component --  the  process causing  state
transitions,  and  the  ``potential''   component  --  the  file  that
maintains  the state.   It  would  be interesting  to  explore for  an
extremum principle for computation.

\newpage
\bibliographystyle{unsrt}
\bibliography{refs}
\end{document}